\documentclass[preprint,aps,showpacs]{revtex4}%
\usepackage{amsfonts}
\usepackage{amsmath}
\usepackage{amssymb}
\usepackage{graphicx}
\usepackage{epsfig}%
\begin{document}
\title{Comment on \textquotedblleft Quantum Key Distribution with Classical
Bob\textquotedblright}
\author{Yong-gang Tan}
\email{ygtan@lynu.edu.cn}
\affiliation{Physics and Information Engineering College, Luoyang Normal College, Luoyang
471022, Henan, People's Republic of China}
\author{Hua Lu}
\affiliation{Department of Physics, Hubei University of Technology, Wuhan 430068, People's
Republic of China}
\author{Qing-yu Cai}
\email{qycai@wipm.ac.cn}
\affiliation{State Key Lab of Magnetics Resonance and Atom and Molecular Physics, Wuhan
Institute of Physics and Mathematics Wuhan 430071, People's Republic of China }

\maketitle

M. Boyer \emph{et al.} \cite{Boyer07} recently proposed an
interesting quantum key distribution scheme (BKM07). It claimed that
Bob doesn't need quantum capacity to ensure the protocol's security.
That is to say, a "classical" Bob can ensure the security of the
key. This work is conceptually novel and interesting. However, in
this comment, we will show that classical Bob is not good enough for
detecting a powerful Eve's eavesdropping.

In BKM07, when Alice's photons flying into Bob's Lab, Bob measures
about half of the incoming photons to generate key and reflects back
the others. Among the registered photons on Bob's detectors, Alice
and Bob drop the results prepared on the $X$ basis and keep the left
as their raw key. Then Alice and Bob's photons can be classified
into two categories: the CTRL photons which are reflected back to
Alice and the SIFT photons which are used to generate key. If Eve
has tagged all of Alice's photons before they enter Bob's realm, she
can differentiate Bob's SIFT photons from CTRL photons: Bob consumed
all the SIFT photons during the course of his measurement, so he has
to send fresh photons which are not tagged in the SIFT mode.
Therefore, Eve can distinguish the SIFT photons from the CTRL
photons in the return line and she can thus obtain the information
of the INFO bits by using the method in the \textit{mock protocol}
presented in~\cite{Boyer07}.

In fact, Eve's tag can be finished with practical technology.
Suppose Eve has an optical wavelength converter which can provide a
very small wavelength change to the aim
photons~\cite{interpretation1}. In practice, the information may be
encoded on the photon's polarization (If the phase-encoding was
used, Eve can select to tag the polarization of the travel
photons.). Since the polarization is communicative with the
wavelength, Eve's operation does not affect the information encoded
on the photons. A practical eavesdropping scheme can hence be
depicted as following.

1. Alice prepares a string of photons randomly in the $X$ basis or
in the $Z$ basis. Let the wavelength of Alice's photons be
$\lambda$.

2. Eve performs a CNOT from the incoming photons into a
$|0\rangle_{blank}$ ancilla before they entering the wavelength
converter. The wavelength becomes $\lambda+\delta\lambda$ after the
photons passed through Eve's Lab. Eve forwards the tagged photons to
Bob.

3. Bob randomly operates the incoming photons in the CTRL mode or in the SIFT
mode. In the former case, Bob just reflects Alice's photons back to Alice. In
the latter case, he measures the photons in the $Z$ basis to read out the
information, then he makes a copy of the information he obtained on a string
of fresh photons and sends them to Alice.

4a. Eve operates a CNOT conversely on the tagged photons as that in step2
which can reset her ancilla and erase the interaction on the initial photon
prepared by Alice.

4b. If Bob's photons are not tagged, Eve measure her ancilla on the $Z$ basis
to extract its information. It is the same as the information Bob read from
Alice's photons.

5. Alice and Bob declare which mode the photons are operated in. If
the quantum bit error rate (QBER) is below a tolerant threshold,
they will use the information obtained from the $Z$-SIFT mode as
their raw key. Or else, they discard the protocol.

In Eve's eavesdropping, if $\delta\lambda$ is chosen appropriately,
the tagged photons can register on Bob's detectors correctly. With
the above eavesdropping scheme, Eve may obtain all Alice and Bob's
information without being detected.

Thus we have showed the classical Bob is not good enough to discover a
powerful Eve. Furthermore, the practical apparatuses of Alice and Bob can not
be the same. The photons prepared by Bob may have different characters with
that of Alice's and then a powerful Eve can distinguish Alice's photons from
that of Bob's. In this case, Eve even doesn't need to tag Alice's photon but
Bob himself tags the travel photons.

This work is funded by NSFC under Grant No. 10504039 and Wuhan
Chenguang Project. Y.-G. Tan also thank the youth fund of Luoyang
Normal College.

\end{document}